\documentclass[twocolumn,showpacs,amsmath,amstex,amssymb,prb]{revtex4}
\usepackage{graphicx}

\newcommand{\ca}{{\bf A}}

\newcommand{\br}{{\bf r}}

\begin{document}
\author{E. Prodan and R. Car}
\address{Department of Chemistry and Princeton Institute fot the Science and Technology of Materials, Princeton University, Princeton, NJ 08544}
\title{DC Conductance of Molecular Wires}

\begin{abstract}
Inspired by the work of Kamenev and Kohn, we present a general discussion of the two-terminal dc conductance of molecular devices within the framework of Time Dependent Current-Density Functional Theory. We derive a formally exact expression for the adiabatic conductance and we discuss the dynamical corrections. For junctions made of long molecular chains that can be either metallic or insulating, we derive the exact asymptotic behavior of the adiabatic conductance as a function of the chain's length. Our results follow from the analytic structure of the bands of a periodic molecular chain and a compact expression for the Green's functions. In the case of an insulating chain, not only do we obtain the exponentially decaying factors, but also the corresponding amplitudes, 
which depend very sensitively on the electronic properties of the contacts. We illustrate the theory by a numerical study of a simple insulating structure connected to two metallic jellium leads. 

\end{abstract}

\date{\today}

\maketitle

\section{Introduction}

The theory of charge transport is often revisited these days, half century after  the first rigorous quantum formulation of transport by Kohn and Luttinger,\cite{Kohn:1957uq, Kohn:1958kx} and after the seminal work of Landauer.\cite{Landauer:1957fk} The on going interest is, of course, propelled by the recent advances in transport measurements at the molecular scale, where all the elements that make this problem extremely difficult, from the theoretical and experimental point of view, become equally important. The challenge comes from the fact that we are dealing with out-of-equilibrium, open, interacting, quantum, many-electron systems. Some recent progress in the field includes the derivation of the Landauer formulas by Kamenev and Kohn using a linear response (as opposed to a scattering) approach,\cite{A.-Kamenev:2001uq} the discovery of a previously unknown dynamical correction to the resistance by Sai and co-workers, the development of a master equation approach to transport by Burke, Car and Gebauer,\cite{Burke:2005kx} the elucidation of the role of non local exchange and correlation effects on transport by Burke and Koentopp,\cite{Koentopp:2006ys} the recent discussion of the non-equilibrium currents by Doyon and Andrei,\cite{Doyon:2006fk} and the exact solution for transport through a quantum dot given by Mehta and Andrei.\cite{Mehta:2006uq} All these results remind us that quantum charge transport is still a work in progress, even at the very fundamental level.

The recent work of Kamenev and Kohn\cite{A.-Kamenev:2001uq} introduced an alternative approach to the scattering method for the calculation of the two- and four-terminal conductance of molecular devices. Although developed within the Hartree approximation, the Kamenev-Kohn approach can be straightforwardly implemented within the  framework of Time Dependent Current-Density Functional Theory.\cite{G.-Vignale:1996fk} We complete this step in the first part of the paper, where we show that it is possible to obtain a formally exact expression for the adiabatic two-terminal dc conductance. In addition, we discuss the dynamical contributions to the conductance. When the transverse currents are neglected, we recover the dynamical correction to the adiabatic conductance discussed by Sai and coworkers.\cite{Sai:2005uq} We argue that the transverse currents may add further dynamical corrections to the conductance.

The rest of the paper focuses on the adiabatic two-terminal dc conductance. We are interested in the transport characteristics of long atomic/molecular chains, either metallic or insulating, attached to metallic leads. Our goal is to derive the exact asymptotic behavior of the conductance with the chain's length. This study complements the analysis of Kamenev and Kohn, who focused on short junctions. The technique that we use in this paper was previously employed in Ref.~\onlinecite{Prodan:2005vn} to study the asymptotic behavior of various perturbations of the electron density in metals and insulators, and it is based on the analytic structure of the bands and a compact expression for the Green's function, as discussed in Ref.~\onlinecite{Prodan:2006yq}. In the present paper, we add a new interesting object, a generalized Wronskian with some special properties that is very useful in evaluating the conductivity tensor of our systems.

For insulating chains, we obtain the usual exponential decay behavior of the conductance with the length of the chains. Depending on the structure of the valence and conduction bands of the chain, we find that the exponential behavior can be associated to more than one relevant exponentially decaying term. Carbon nanotubes are typical examples of a system in which several exponentially decaying terms are important.\cite{Pomorski:2004nx} We find that the asymptotic behavior of the conductance with the chain's length is not just exponentially decaying, but may contain modulating factors. The exponential decay constants and the period of the oscillations can be predicted from complex band structure calculations. In addition, we also obtain the exact expression for the amplitudes of the exponentially decaying terms. 

Our work makes a rigorous connection between complex band structure and tunneling. This connection was made in Ref.~\onlinecite{Mavropoulos:2000cr} and, since then, it was further discussed in several theoretical and computational studies, some of which are mentioned later in our paper. However, although the existence of such connection is now quite obvious, rigorous and explicit expressions for the amplitudes of the different exponentially decaying terms were missing. Ref.~\onlinecite{Mavropoulos:2000cr} proposes, for example, that the amplitudes are directly proportional to the number of states at the complex $k$ vector. This is only partially true. In Ref.~\onlinecite{Tomfohr:2004ve}, the amplitude is simply set to unity without supporting arguments. The explicit expressions for the amplitudes, given in the present paper, show that they are determined by the overlap integral between the spectral kernel, the complex $k$ Bloch functions of the chain and a suitably defined potential. The amplitudes are determined by the electronic properties in the immediate vicinity of the contacts. The expressions are intuitive and simple enough to allow for estimates, without the need for costly numerical calculations. 

We also derive explicit expressions for the adiabatic conductance of metallic chains, which show the well known oscillatory behavior with the chain's length.\cite{Lang:1997zr,Lang:1998vn,Smit:2003uq,Lee:2004kx,Khomyakov:2006ys} We discuss the phase and the wavelength of the oscillation in the case of monovalent atom chains in the light of our analytic results.

In the last section of the paper, we present a numerical implementation of these ideas to a simple insulating chain. We illustrate all the elements entering our expressions for the conductance. For example, we compute and display the Riemann surface of the bands. 

\section{Conductance}

\begin{figure}
  \includegraphics[width=4cm]{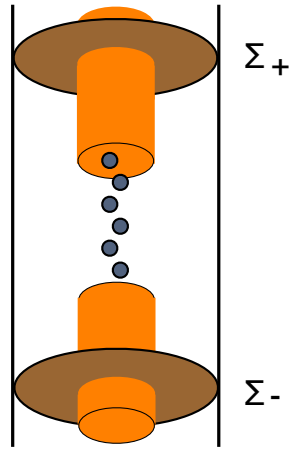}\\
  \caption{The geometry of the chain+leads structure.}
  \label{geometry}
\end{figure}

We consider a charge transport experiment involving a molecular chain attached to metallic leads, as shown in Fig.~\ref{geometry}. Following Kamenev and Kohn,\cite{A.-Kamenev:2001uq} we derive a general and formally exact expression for the two-terminal dc conductance of such system.

We start from the Vignale-Kohn linear response equation,\cite{G.-Vignale:1996fk}
\begin{equation}\label{VignaleKohn}
{\bf j}(\br,\omega)=\int \hat{\chi}^{\text{\tiny{KS}}}(\br,\br';\omega) \ca^{\text{eff}}_1(\br',\omega)d\br',
\end{equation}
which gives the expectation value of the current when the system is coupled to a time oscillating electromagnetic field. As in Ref.~\onlinecite{A.-Kamenev:2001uq}, we obtain the dc regime by letting $\omega$ go to zero. 

In Eq.~(\ref{VignaleKohn}), $\hat{\chi}^{\text{\tiny{KS}}}(\br,\br';\omega)$ is the equilibrium Kohn-Sham current-current correlation tensor and $\ca^{\text{eff}}_1(\br,\omega)$ is an effective vector potential. Given the particular gauge choice, we can define an effective electric field as ${\bf E}^{\text{\tiny{eff}}}_1=\partial_t \ca^{\text{eff}}_1$,\cite{G.-Vignale:1996fk} in which case the linear response equation becomes
\begin{equation}\label{dcurrent}
{\bf j}(\br,\omega)=\int \hat{\sigma}^{\text{\tiny{KS}}}(\br,\br';\omega) {\bf E}^{\text{\tiny{eff}}}_1(\br',\omega)d\br'.
\end{equation}
A simple and explicit expression for ${\bf E}^{\text{\tiny{eff}}}_1(\br',\omega)$ is given in Ref.~\onlinecite{Vignale:1997fk}:
\begin{equation}\label{Eeff}
{\bf E}^{\text{eff}}_1= \frac{1}{e}\nabla \phi_1^{\text{\tiny{ext}}}+\frac{1}{e}\nabla \phi^{\text{\tiny{HXC}}}_1+{\bf E}^{\text{\tiny{dyn}}}_1,
\end{equation}
where $\phi^{\text{\tiny{HXC}}}_1$ is the adiabatic contribution, i.e. the linearized static Hartree-exchange-correlation potential, and ${\bf E}^{\text{\tiny{dyn}}}_1$ is the dynamical part of ${\bf E}^{\text{\tiny{eff}}}_1$, given by
\begin{equation}
{\bf E}^{\text{dyn}}_1=-\frac{1}{en_0}\nabla \hat{\zeta}.
\label{Edyn}
\end{equation}
Here, $\hat{\zeta}$ is the viscoelastic stress tensor.

$\hat{\sigma}^{\text{\tiny{KS}}}$ in Eq.~(\ref{dcurrent}) is the Kohn-Sham conductivity tensor. In the limit $\omega \rightarrow 0$, the conductivity tensor reduces to
\begin{equation}
\sigma_{\alpha \beta}^{\text{\tiny{KS}}}(\br,\br')=\frac{1}{2\pi}\text{Tr} \left \{ \hat{j}_\alpha(\br) \ G_{\epsilon_F^+}^{\text{\tiny{KS}}}\ \hat{j}_\beta(\br') \ G_{\epsilon_F^-}^{\text{\tiny{KS}}} \right \},
\end{equation}
where $\epsilon_F^\pm=\epsilon_F \pm i\delta$, $G_\epsilon^{\text{\tiny{KS}}}$ is the Green's function of the equilibrium Kohn-Sham system,
\begin{equation}
G_\epsilon^{\text{\tiny{KS}}}=(\epsilon-H_{\text{\tiny{KS}}})^{-1},
\end{equation}
and $\hat{\bf j}$ is the current operator. A convenient expression for $\hat{\sigma}$ is
\begin{equation}\label{basicsigma}
\sigma_{\alpha \beta}^{\text{\tiny{KS}}}(\br,\br')=\frac{1}{4\pi} 
G_{\epsilon_F^+}^{\text{\tiny{KS}}}(\br,\br') \ \overleftrightarrow{\partial_\alpha} \ 
\overleftrightarrow{\partial'_\beta} \ G_{\epsilon_F^-}^{\text{\tiny{KS}}}(\br',\br),
\end{equation}
where we used the shorthand $\overleftrightarrow{\partial_\alpha} =\overrightarrow{\partial}_\alpha-\overleftarrow{\partial}_\alpha$.
An important property of the Kohn-Sham conductivity at zero frequency is
\begin{equation}\label{property}
\sum_\alpha \partial_\alpha \sigma_{\alpha \beta}^{\text{\tiny{KS}}}(\br,\br') =\sum_\beta \partial'_\beta \sigma_{\alpha \beta}^{\text{\tiny{KS}}}(\br,\br')=0,
\end{equation}
which follows either from the continuity equation applied to Eq.~(\ref{dcurrent}) or directly from Eq.~(\ref{basicsigma}).

Because of the spatial confinement, the Kohn-Sham conductivity tensor goes rapidly to zero as one moves laterally away from the chain-leads structure. We can then consider the chain+leads system inside a tube that is large enough that the conductivity tensor is practically zero at the tube surface and beyond. The net current flowing through the molecular chain is given by
\begin{equation}\label{netcurrent}
I=\int_{\Sigma} d{\bf S} \int d\br'  \ \hat{\sigma}^{\text{\tiny{KS}}}(\br,\br') {\bf E}^{\text{eff}}_1(\br'),
\end{equation}
where $\Sigma$ is an arbitrary transversal section and the integral over $\br'$ is taken only inside the tube. 
We break the current in Eq.~(\ref{netcurrent}) as $I = I^{\text{\tiny{ad}}} + I^{\text{\tiny{dyn}}}$, where $I^{\text{\tiny{ad}}}$ is the current resulting only from the adiabatic part of the effective electric field, ${\bf E}_1 ^{\text{\tiny{ad}}}= \nabla(\phi_1^{\text{\tiny{ext}}}+\phi_1^{\text{\tiny{HXC}}})\equiv \nabla \phi_1^{\text{\tiny{ad}}}$, and $I^{\text{\tiny{dyn}}}$ is the current resulting from the dynamical part of ${\bf E}^{\text{eff}}_1$.

To get a clean expression for the conductance, one needs to pull out of the integral the physical electric potential drop between points at $z=\pm \infty$. This would be straightforward if one could make the simplifying assumption that the effective electric field is uniform in the lateral direction. This is, however, a gross approximation for the structure in Fig.~\ref{geometry}. The difficulty is not present in Ref.~\onlinecite{Bokes:2004zr}, which considers a different linear response equation,   involving the full many-body conductivity tensor and the external field. Since one has control on the external field, it can be considered uniform in the lateral direction, greatly simplifying the issue.

Let us first define a conductance for $I^{\text{\tiny{ad}}}$ and then comment on the xc contributions. $I^{\text{\tiny{ad}}}$ is given by:
\begin{equation}\label{netcurrentp}
I^{\text{\tiny{ad}}}=\int_{\Sigma} d{\bf S} \int d\br'  \ \hat{\sigma}^{\text{\tiny{KS}}}(\br,\br') \nabla \phi_1^{\text{\tiny{ad}}}(\br').
\end{equation}
Due to Eq.~(\ref{property}),
\begin{equation}
\hat{\sigma}^{\text{\tiny{KS}}}(\br,\br')\nabla'\phi^{\text{eff}}_1(\br')=\nabla' \hat{\sigma}^{\text{\tiny{KS}}}(\br,\br')\phi_1(\br'),
\end{equation}
which allows us to transform the volume integral over $\br'$ in Eq.~(\ref{netcurrentp}) into a surface integral. First, we consider this integral over a finite volume, between the $\Sigma_\pm$ surfaces of Fig.~\ref{geometry}, and then take the infinite volume limit by moving the surfaces at $z=\pm \infty$. An integration by parts in Eq.~(\ref{netcurrent}) gives
\begin{equation}
I^{\text{\tiny{ad}}}=\int_{\Sigma} d S_\alpha \left ( \int_{\Sigma_+} -\int_{\Sigma_-}  \right ) d S'_\beta \ \sigma_{\alpha \beta}^{\text{\tiny{KS}}}(\br,\br') \phi_1^{\text{\tiny{ad}}}(\br').
\end{equation}
Next we deform the sections $\Sigma_\pm$ into surfaces of constant potential,
\begin{equation}
\phi_1^{\text{\tiny{ad}}}(\br)\vert_{\Sigma_\pm}=\phi_\pm^{\text{\tiny{ad}}}.
\end{equation}
This is possible because $\Sigma_\pm$ are arbitrary sections, which are used here only to take the infinite volume limit. Then it follows that
\begin{eqnarray}
I^{\text{\tiny{ad}}}=\phi_+^{\text{\tiny{ad}}} \int_{\Sigma} d S_\alpha   \int_{\Sigma_+} d S'_\beta \ \sigma_{\alpha \beta}^{\text{\tiny{KS}}}(\br,\br') \\
-\phi_-^{\text{\tiny{ad}}} \int_{\Sigma} d S_\alpha \int_{\Sigma_-}  d S'_\beta \ \sigma_{\alpha \beta}^{\text{\tiny{KS}}}(\br,\br'). \nonumber
\end{eqnarray}
After pulling the potential out,  the integrals become independent of the surfaces, due to Eq.~(\ref{property}). Therefore, if we deform the sections so that $\Sigma$ lies on the $xy$ plane at some arbitrary $z$ and $\Sigma_\pm$ on the $xy$ plane at some arbitrary $z'$ and take the infinite volume limit, we finally obtain
\begin{equation}
I^{\text{\tiny{ad}}}=\Delta \phi_{\infty}^{\text{\tiny{ad}}} \int d \br_\bot  \int d\br'_\bot \ \sigma_{zz}^{\text{\tiny{KS}}}(\br_\bot,z,\br'_\bot, z').
\end{equation}
It was argued in Ref.~\onlinecite{Koentopp:2006ys} that, within the commonly used density functionals, the xc contribution to $\Delta \phi_{\infty}^{\text{\tiny{ad}}}$ is identically zero. The argument follows from the observation that
 \begin{equation}\label{Rad}
\Delta \phi_{\infty}^{\text{\tiny{xc}}}= \left. \frac{\delta v_{\text{xc}}}{\delta n} n_1\right|_{z=+\infty}-\left. \frac{\delta v_{\text{xc}}}{\delta n} n_1\right|_{z=-\infty}
 \end{equation}
 and the fact that the perturbed density $n_1$ is localized near the junction. Thus we conclude that  $\Delta \phi_{\infty}^{\text{\tiny{ad}}}$ is in fact the physical electric potential drop $\Delta \phi_\infty$ and consequently $I^{\text{\tiny{ad}}}= g_0 \Delta \phi_\infty$, where $g_0$ is a constant depending solely on the equilibrium properties of the the system:
\begin{equation}\label{adiabaticg}
g_0 \equiv \int d \br_\bot  \int d\br'_\bot \ \sigma_{zz}^{\text{\tiny{KS}}}(\br_\bot,z,\br'_\bot, z').
\end{equation}
This also shows that the conductance derived from the adiabatic approximation of the time dependent xc potential is exactly the $g_0$ of Eq~(\ref{adiabaticg}). Thereafter, we call $g_0$ the adiabatic conductance. Note that the expression for $g_0$ is formally identical to the one derived by Kamenev and Kohn within the Hartree approximation,\cite{A.-Kamenev:2001uq} or the one derived by Baranger and Stone for non-interacting electrons.\cite{Baranger:1989bs}

For the net current, we can conclude at this point that $I = (g_0+I^{\text{\tiny{dyn}}} / \Delta \phi_\infty)\Delta \phi_\infty$, so we can write the total conductance $g$ as the formal sum $g=g_0+g^{\text{\tiny{dyn}}}$, where $g^{\text{\tiny{dyn}}} \equiv I^{\text{\tiny{dyn}}}/\Delta \phi_\infty$. If we {\it neglect} the transversal part of ${\bf E}_1^{\text{\tiny{dyn}}}$ and write it as the gradient of a dynamic potential $\phi_1^ {\text{\tiny{dyn}}}$, we can follow the same steps as we did for $I^{\text{\tiny{ad}}}$ and prove that $I^{\text{\tiny{dyn}}}=g_0\Delta^{\text{\tiny{dyn}}}_\infty$. In other words, 
\begin{equation}
I = g_0(\Delta \phi_\infty + \Delta \phi_\infty ^{\text{\tiny{dyn}}}),
\end{equation}
which is precisely one of the key equations (see Eq.~12) in Ref.~\onlinecite{Sai:2005uq}. We are then led to conclude that there is an additional dynamical correction to the one discussed in this reference, correction that comes from the transverse part of ${\bf E}_1^{\text{\tiny{dyn}}}$.

To evaluate $g^{\text{\tiny{dyn}}}$, one needs to solve the self-consistent equation~(\ref{dcurrent}) for the current and then compute $I^{\text{\tiny{dyn}}}$. It seems unlikely that one could carry analytic work on $g^{\text{\tiny{dyn}}}$ beyond that of Ref.~\onlinecite{Sai:2005uq}. We succeed, however, in deriving analytic expressions for $g_0$ in several cases that complement the work of Kamenev and Kohn.\cite{A.-Kamenev:2001uq}

\section{Strictly one dimensional systems}

In order to formulate a strategy for calculating the adiabatic conductance, it is useful to start with the simple case of a strictly one dimensional system. In this case, the expression for $g_0$ simplifies to:
\begin{equation}\label{1dg}
g_0 =\frac{1}{4\pi} G_{\epsilon_F^+}(x,x')\overleftrightarrow{\partial_x} \overleftrightarrow{\partial_{x'}} G_{\epsilon_F^-}(x,x').
\end{equation}
The Green's function can always be expanded using eigenvectors and eigenvalues. However, the large number of terms generated by such expansions are hard to control analytically. Our aim is to express $g_0$ in terms of a small number of parameters with straightforward and intuitive physical meaning. For this, we use a compact expression for the Green's function,
\begin{equation}\label{compactGreen}
G_\epsilon(x,x')=2\frac{\psi_\epsilon^<(x_<)\psi_\epsilon^>(x_>)}{W(\psi_\epsilon^<,\psi_\epsilon^>)},
\end{equation}  
where $x_{<} / x_{>}=\min/\max(x,x')$ and $\psi_\epsilon^{</>}(x)$ are the solutions of the Schroedinger equation at energy $\epsilon$, satisfying the boundary condition either to the right or to the left. For our infinite system, these boundary conditions are simply $\psi_\epsilon^{</>}(x) \rightarrow 0$ as $x \rightarrow \mp \infty$, respectively. We always evaluate the Green's function at an energy $\epsilon$ outside the allowed energy spectrum or, at most, take the limit of the Green's function for $\epsilon$  approaching the energy spectrum. It follows from the standard theory of second order differential equations in 1D that the solutions $\psi_\epsilon^\gtrless$ are uniquely defined by these boundary conditions. 

One clarification is needed here. When we talk about a solution of the Schroedinger equation we do not mean an eigenfunction. An eigenfunction is a solution satisfying simultaneously the boundary conditions to the left {\it and} to the right. While the Schroedinger equation has solutions at any energy, eigenfunctions exist only for certain energies which define the energy spectrum.

$W(\psi,\phi)$ denotes the Wronskian of $\psi$ and $\phi$, i.e.
\begin{equation}\label{wronskian}
W(\psi,\phi)=\psi(x)\overleftrightarrow{\partial_x}\phi(x).
\end{equation}
If $\psi$ and $\phi$ are two solutions of the Schroedinger equation at the same energy, the right hand side of Eq.~(\ref{wronskian}) does not depend on $x$. Using the Wronskian, we can rewrite the adiabatic conductance as:
\begin{equation}\label{finalg}
g_0=\frac{1}{\pi}\frac{W(\psi^>_{\epsilon_F^+},\psi^>_{\epsilon_F^-})W(\psi^<_{\epsilon_F^+},\psi^<_{\epsilon_F^-})}{W(\psi_{\epsilon_F^+}^<,\psi_{\epsilon_F^+}^>)W(\psi_{\epsilon_F^-}^<,\psi_{\epsilon_F^-}^>)}.
\end{equation}
In the next subsections we evaluate this expression for several cases of interest.

\begin{figure}
  \includegraphics[width=8.6cm]{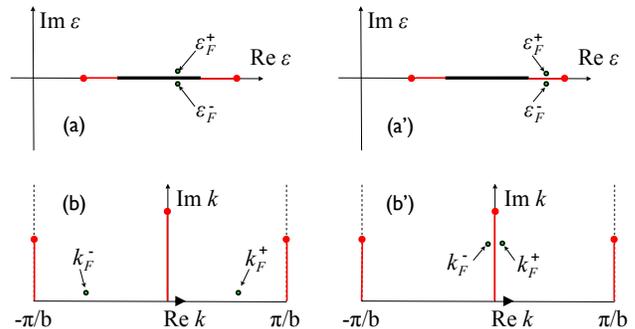}\\
  \caption{(a and a') The position of $\epsilon_F^\pm$ in the complex energy plane and (b and b') the corresponding $k_F^\pm$ in the complex $k$-plane for metals (left) and insulators (right). The figure also shows the two branch points (red dots), one at $k=0$ and one at the zone boundary, and their corresponding energies.}
 \label{RS}
\end{figure} 

\subsection{Small junction and clean, long metallic leads} 

For simplicity we assume that the junction perturbs the periodic potential of the clean leads only inside a finite interval $[0,L]$. We denote by $b$ the lattice constant and by $\psi_k(x)$ the Bloch functions of the leads. The wavenumber $k$ is complex in general, except when $\epsilon$ is on the real axis, inside the allowed energy bands. As a convention, $k$ will always be chosen with positive imaginary part ($-k$ will then always have negative imaginary part). In 1D and only in 1D, for each energy $\epsilon$ (not necessarily real), there are two and only two independent Bloch solutions, $\psi_k(x)$ and $\psi_{-k}(x)$.   The Bloch solutions are normalized according to
\begin{equation}
\frac{1}{b}\int_0^b \psi_k(x) \psi_{-k}(x) dx =1,
\end{equation}
and their phase is fixed by requiring 
\begin{equation}
\psi_k(0)=\psi_{-k}(0).
\end{equation}
One can use the fundamental property of the Bloch solutions, 
\begin{equation}\label{fproperty}
\psi_k(x+b)=e^{ikb}\psi_k(x),
\end{equation}
to test their asymptotic behavior. One finds that $\psi_k(x)$ decays to zero when $x \rightarrow \infty$ while it diverges when $x\rightarrow -\infty$ and $\psi_{-k}(x)$ decays to zero when $x \rightarrow -\infty$ while it diverges when $x\rightarrow \infty$. There is an exception to this behavior, namely when the energy is in the allowed energy bands. In this case, the Bloch functions behave like waves. Instead of using band indices, it is more convenient to think that the wavenumber $k$ lies on a Riemann surface.\cite{Kohn:1959fk}

When the clean leads are perturbed by the junction, we have:
\begin{equation}\label{sol1}
\psi_{\epsilon}^>(x)= \left \{ \begin{array}{ll}
\psi_{k}(x)+{\cal R}^-(k)\psi_{-k}(x), & x<-L \medskip\\
{\cal T}(k) \psi_{k}(x), & x>0
\end{array} \right.
\end{equation}
and
\begin{equation}\label{sol2}
\psi_{\epsilon}^<(x)= \left \{ \begin{array}{ll}
{\cal T}(k) \psi_{-k}(x), & x<-L \medskip\\
\psi_{-k}(x)+{\cal R}^+(k)\psi_{k}(x), & x>0\\
\end{array} \right.
\end{equation}
where $k$ is the unique wavenumber with Im($k$)$>$0 such that $\epsilon_k$=$\epsilon$. ${\cal T}(k)$ and ${\cal R}^\pm(k)$ in Eqs.~\ref{sol1}-\ref{sol2} are the transmission and reflection coefficients of the junction.

To compute the conductance (Eq.~\ref{finalg}) we need the solutions \ref{sol1}-\ref{sol2} at $\epsilon_F^\pm$. $\epsilon_F$ is located inside an allowed band of the leads and $\epsilon_F^\pm$ is immediately above/below the allowed band, as shown in panel (a) of Fig.~\ref{RS}. The corresponding wavenumbers $k_F^\pm$ are shown in panel (b) of the same figure. In the limit $\delta \rightarrow 0$, $k_F^+=-k_F^-=k_F$, we obtain:
\begin{equation}
\psi_{\epsilon_F^\pm}^>(x)= \left \{ \begin{array}{ll}
\psi_{\pm k_F}(x)+{\cal R}^-(\pm k_F)\psi_{\mp k_F}(x), & x<-L \medskip\\
{\cal T}(\pm k_F) \psi_{\pm k_F}(x), & x>0
\end{array} \right.
\end{equation}
and
\begin{equation}
\psi_{\epsilon_F^\pm}^<(x)= \left \{ \begin{array}{ll}
{\cal T}(\pm k_F) \psi_{\mp k_F}(x), & x<-L \medskip\\
\psi_{\mp k_F}(x)+{\cal R}^+(\pm k_F)\psi_{\pm k_F}(x), & x>0\\
\end{array} \right.
\end{equation}

By taking $x$ in the right lead, and  noticing that ${\cal T}(-k)={\cal T}(k)^*$, we can easily compute a first set of Wronskians:
\begin{equation}
\begin{array}{l}
W(\psi^>_{\epsilon_F^+},\psi^>_{\epsilon_F^-})=|{\cal T}(k_F)|^2 W_0 \medskip
\\
W(\psi^<_{\epsilon_F^+},\psi^>_{\epsilon_F^+})=-{\cal T}(k_F)W_0.
\end{array}
\end{equation}
By taking $x$ in the left lead, we can calculate the remaining Wronskians appearing in Eq.~(\ref{finalg}):
\begin{equation}
\begin{array}{l}
W(\psi^<_{\epsilon_F^+},\psi^<_{\epsilon_F^-})=-|{\cal T}(k_F)|^2 W_0 \medskip
\\
W(\psi^<_{\epsilon_F^-},\psi^>_{\epsilon_F^-})={\cal T}(-k_F)W_0,
\end{array}
\end{equation}
where $W_0=W(\psi_{k_F},\psi_{-k_F})$. After straightforward cancellations,  Eq.~(\ref{finalg}) leads to the classic result:\cite{Landauer:1957fk}
\begin{equation}
g_0=\frac{1}{\pi}|{\cal T}(k_F)|^2.
\end{equation}

\subsection{Long chains} 

Here we consider the opposite case of a clean, long molecular or atomic chain attached to arbitrary metallic leads. This case is especially important for understanding the experiments on self-assembled monolayers (SAM) grown on a clean substrate. In these experiments the conductance is probed by an STM tip or a Hg droplet placed above the SAM.\cite{Cui:2002dq,Wold:2001bh,A.-Salomon:2005kx,Nesher:2006ys} In this setup, the left and right ``leads" are very different and the picture in which the junction acts as a scattering center for the states of the leads is no longer appropriate.

\begin{figure}
  \includegraphics[width=7cm]{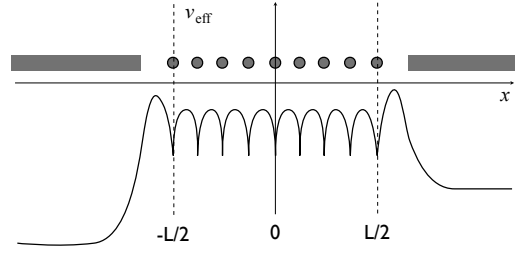}\\
  \caption{Schematic of the effective potential of the 1D chain.}
  \label{1dchain}
\end{figure}

Deferring discussion of a realistic effective potential to the following sections, we assume here a long but finite molecular chain attached to metallic leads that is described by an effective potential shown, qualitatively, in Fig.~\ref{1dchain}. The main simplification is that, inside the interval $[-L/2,L/2]$, the effective potential is strictly periodic.

By extending this periodic potential to $\mp \infty$, one can calculate the corresponding energy bands. If the Fermi energy of the system, i.e of the infinite leads plus the finite molecular chain, falls inside an allowed band we call the corresponding molecular chain metallic. We call it insulating otherwise.
  
Exploiting the fact that Eq.~(\ref{1dg}) is independent of $x$ and $x'$, we conveniently fix $x$ and $x'$ deep inside the chain. Then we can use the Bloch solutions of the molecular chain in a similar way to what we did in the previous subsection with the Bloch solutions of the leads. Normalization and phase of the Bloch solutions are fixed in the same way.
 
Inside the $[-L/2,L/2]$ interval, the exact solutions of the Schroedinger equation are
\begin{equation}\label{gensolutions}
\begin{array}{l}
\psi^>_{\epsilon}(x)=\psi_{k}(x)+{\cal R}_\text{\tiny{R}}(k) \psi_{-k}(x) \medskip
\\
\psi^<_{\epsilon}(x)=\psi_{-k}(x)+{\cal R}_\text{\tiny{L}} (k) \psi_{k}(x),
\end{array}
\end{equation}
where $k$ is again the unique wave number in the upper complex semi-plane such that $\epsilon_k=\epsilon$. ${\cal R}_\text{\tiny{L/R}}(k)$ are the reflection coefficients of the left/right contacts.
The solutions formally look the same for metallic and insulating chains, but the meaning of $k_F^\pm$ and the behavior of the reflection coefficients at these wavenumbers are different in the two cases.

{\it Metallic chains.} In this case $\epsilon_F$ is inside an allowed band and we can follow the notation in panels (a) and (b) of Fig.~\ref{RS}. Taking the limits $\epsilon\rightarrow \epsilon_F^\pm$ and $\delta \rightarrow 0$ we have:
\begin{equation}
\begin{array}{l}
\psi^>_{\epsilon_F^\pm}(x)=\psi_{\pm k_F}(x)+{\cal R}_\text{\tiny{R}}(\pm k_F) \psi_{\mp k_F}(x) \medskip
\\
\psi^<_{\epsilon_F^\pm}(x)=\psi_{\mp k_F}(x)+{\cal R}_\text{\tiny{L}} (\pm k_F) \psi_{\pm k_F}(x),
\end{array}
\end{equation}
leading to:
\begin{equation}
\begin{array}{l}
W(\psi^>_{\epsilon_F^+},\psi^>_{\epsilon_F^-})=[1-|{\cal R}_\text{\tiny{R}}(k_F)|^2] W_0 \medskip
\\
W(\psi^<_{\epsilon_F^+},\psi^<_{\epsilon_F^-})=-[1-|{\cal R}_\text{\tiny{L}}(k_F)|^2] W_0 
\end{array}
\end{equation}
and
\begin{equation}
\begin{array}{l}
W(\psi^<_{\epsilon_F^+},\psi^>_{\epsilon_F^+})=-[1-{\cal R}_\text{\tiny{L}}(k_F){\cal R}_\text{\tiny{R}}(k_F)] W_0 \medskip
\\
W(\psi^<_{\epsilon_F^-},\psi^>_{\epsilon_F^-})=[1-{\cal R}_\text{\tiny{L}}(-k_F){\cal R}_\text{\tiny{R}}(-k_F)] W_0.
\end{array}
\end{equation}
Since ${\cal R}_\text{\tiny{L/R}}(-k_F)={\cal R}_\text{\tiny{L/R}}(k_F)^*$, we obtain the following simple expression for the adiabatic conductance:
\begin{equation}\label{1dmetalg}
g_0= \frac{1}{\pi}\frac{[1-|{\cal R}_\text{\tiny{L}}(k_F)|^2] [1-|{\cal R}_\text{\tiny{R}}(k_F)|^2] }{|1-{\cal R}_\text{\tiny{L}}(k_F){\cal R}_\text{\tiny{R}}(k_F)|^2}.
\end{equation}
The reflection coefficients contain a phase factor that depends on the length of the chain and is related to our choice for fixing the phases of the Bloch functions. In Eq.~(\ref{1dmetalg}) we have assumed the phase of the Bloch functions to be zero at $x=0$. If we move the origin at the left/right contacts, the reflection coefficients become independent of L. This requires a rescaling:\cite{Prodan:2006uq}
\begin{equation}\label{rescale}
\begin{array}{c}
{\cal R}_\text{\tiny{L}}(k_F)\rightarrow e^{ik_F L}{\cal R}_\text{\tiny{L}}(k_F) \medskip \\
{\cal R}_\text{\tiny{R}}(k_F)\rightarrow e^{ik_F L}{\cal R}_\text{\tiny{R}}(k_F).
\end{array}
\end{equation}
The expression for the adiabatic conductance becomes
\begin{equation}\label{1dmetalgfinal}
g_0(L)= \frac{1}{\pi}\frac{[1-|{\cal R}_\text{\tiny{L}}(k_F)|^2] [1-|{\cal R}_\text{\tiny{R}}(k_F)|^2] }{|1-e^{2ik_FL} {\cal R}_\text{\tiny{L}}(k_F){\cal R}_\text{\tiny{R}}(k_F)|^2}.
\end{equation}

Eq.~(\ref{1dmetalgfinal}) gives the exact behavior of $g_0$ as a function of $L$. Since this expression applies to arbitrary periodic potentials and arbitrary lead potentials, we conclude that the behavior shown in Eq.~(\ref{1dmetalgfinal}) is universal for 1D metallic chains. The reflection coefficients are numbers between 0 and 1. For poor/good contacts, the magnitude of the reflection coefficients is close to 1/0, respectively.

Resonant transport can also be easily understood from Eq.~(\ref{1dmetalgfinal}). Indeed, $g_0(L)$ is large whenever the denominator is small, which may happen at specific values of $k_F$. Eq.~(\ref{1dmetalg}) provides not only these values, but also the width of the resonance. Indeed, if we look at $g_0$ as a function of $k_F$, we can see that it has poles whenever $k_F$ is equal to a solution of the following equation in $k$:
\begin{equation}\label{poles}
e^{2ikL} {\cal R}_\text{\tiny{L}}(k){\cal R}_\text{\tiny{R}}(k)=1.
\end{equation}
This equation has solutions only in the lower complex semi-plane, which can be obtained by analytical continuation of the reflection coefficients in the lower complex semi-plane. The real part of these solutions gives the resonant values of $k_F$ and the imaginary part gives the width of the resonances. We notice that Eq.~(\ref{poles}) determines also the resonances of the Green's function.\cite{Prodan:2006uq} In other words, the positions and the widths of the transport resonances are exactly the same as those appearing in the density of states projected on the chain. To actually observe resonant transport, the width of a resonance must be smaller than the separation between adjacent resonances.

A consequence of Eq.~(\ref{1dmetalgfinal}) is that $g_0$  oscillates with the chain length. Such behavior was reported previously, both experimentally and theoretically.\cite{Lang:1997zr,Lang:1998vn,Smit:2003uq,Lee:2004kx,Khomyakov:2006ys} According to Eq.~(\ref{1dmetalgfinal}), the wavelength of the oscillation is half the Fermi wavelength of the chain. Chains of monovalent atoms, for example, are half filled and $g_0$ is predicted to be different when the chain contains an odd or an even number of atoms:
\begin{equation}\label{gvn}
g_0(L)= \frac{1}{\pi}\frac{[1-|{\cal R}_\text{\tiny{L}}(k_F)|^2] [1-|{\cal R}_\text{\tiny{R}}(k_F)|^2] }{|1-e^{i \pi N} {\cal R}_\text{\tiny{L}}(k_F){\cal R}_\text{\tiny{R}}(k_F)|^2}.
\end{equation}
This is precisely what has been observed. Interestingly, several independent numerical simulations showed that, for alkali-metal chains,\cite{Lang:1997zr, Lee:2004kx,Khomyakov:2006ys} $g_0$ is larger for an odd number of atoms while for noble-metal chains the opposite occurs.\cite{Brandbyge:1999ly,Lee:2004kx} The behavior for noble-metal chains is still under debate since existing experiments do not seem to confirm the prediction,\cite{Smit:2003uq} and a new study finds that alternative scenarios can happen.\cite{Czerner:2006fv} 

On the basis of expression (\ref{1dmetalgfinal}), we can easily understand these phenomena. In the case of alkali chains, the projected density of states on the chain (PDOS) displays sharp resonances.\cite{Lee:2004kx} The PDOS corresponding to each resonance integrates to 2. Thus, for an odd number of atoms, the Fermi level sits on top of a resonance while for an even number of atoms the Fermi level sits between two resonances. According to our discussion of resonant transport, when the Fermi level sits on top of the resoncance, the denominator in Eq.~(\ref{gvn}) is small, leading to a large value of $g_0$. In fact, if the left and right reflection coefficients are the same, $g_0$ takes the maximum allowed value of $1/\pi$ (a.u.), no matter how bad the contacts are. Now assume that we add one more atom to the chain so that we have an even number of atoms. The Fermi level moves between two resonances. The reflection coefficients have slow energy dependence and do not change much whereas the phase factor in front of them changes sign. The conclusion is that now the denominator of Eq.~(\ref{gvn}) takes a maximum value, leading to a minimum value of $g_0$.

Chains of noble-metal atoms are more jellium like.\cite{Lee:2004kx} For good contacts, we expect the reflection coefficients to be small and real, in which case Eq.~(\ref{gvn}) predicts that $g_0$ is larger for an even number of atoms. This is the behavior found in several numerical simulations.\cite{Brandbyge:1999ly,Lee:2004kx} However, the phase of the reflection coefficients may be quite sensitive to the details of the contacts.\cite{Asai:2005ij,Czerner:2006fv} This can explain the different behavior observed in the experiments. 

{\it Insulating chains.} In this case, the Fermi energy falls within an energy gap. Compared to the metallic case, we have the following differences: $k_F^\pm \rightarrow k_F$ as $\delta \rightarrow 0$, but $k_F$ is now complex as shown in panel (b') of Fig.~\ref{RS}. The reflection coefficients have branch cuts along the red lines in Fig.~\ref{RS}. Consequently, ${\cal R}_\text{\tiny{L/R}}(k_F^+)$ are different from ${\cal R}_\text{\tiny{L/R}}(k_F^-)$.\cite{Prodan:2006uq} Thus, when taking the limit for $\delta \rightarrow 0$ in Eq.~(\ref{gensolutions}), we obtain:
\begin{equation}
\begin{array}{l}
\psi^>_{\epsilon_F^\pm}(x)=\psi_{k_F}(x)+{\cal R}_\text{\tiny{R}}(k_F^\pm) \psi_{-k_F}(x) \medskip
\\
\psi^<_{\epsilon_F^\pm}(x)=\psi_{-k_F}(x)+{\cal R}_\text{\tiny{L}} (k_F^\pm) \psi_{k_F}(x).
\end{array}
\end{equation}

Then:
\begin{equation}
\begin{array}{l}
W(\psi^>_{\epsilon_F^+},\psi^>_{\epsilon_F^-})=[{\cal R}_\text{\tiny{R}}(k_F^-)-{\cal R}_\text{\tiny{R}}(k_F^+)] W_0 \medskip
\\
W(\psi^<_{\epsilon_F^+},\psi^<_{\epsilon_F^-})=[{\cal R}_\text{\tiny{L}}(k_F^+)-{\cal R}_\text{\tiny{L}}(k_F^-)] W_0 
\end{array}
\end{equation}
and
\begin{equation}
\begin{array}{l}
W(\psi^<_{\epsilon_F^+},\psi^>_{\epsilon_F^+})=-[1-{\cal R}_\text{\tiny{L}}(k_F^+){\cal R}_\text{\tiny{R}}(k_F^+)] W_0 \medskip
\\
W(\psi^<_{\epsilon_F^-},\psi^>_{\epsilon_F^-})=-[1-{\cal R}_\text{\tiny{L}}(k_F^-){\cal R}_\text{\tiny{R}}(k_F^-)] W_0.
\end{array}
\end{equation}
This leads to
\begin{equation}
g_0= \frac{4}{\pi}\frac{\mbox{Im}[{\cal R}_\text{\tiny{L}}(k_F^+)] \mbox{Im}[{\cal R}_\text{\tiny{R}}(k_F^+)] }{|1- {\cal R}_\text{\tiny{L}}(k_F^+){\cal R}_\text{\tiny{R}}(k_F^+)|^2}.
\end{equation}
Rescaling the reflection coefficients as in  Eq.~(\ref{rescale}) so that they become independent of $L$,  we finally obtain
\begin{equation}
g_0(L)= \frac{4}{\pi}\frac{\mbox{Im}[{\cal R}_\text{\tiny{L}}(k_F^+)] \mbox{Im}[{\cal R}_\text{\tiny{R}}(k_F^+)] e^{-2\beta L}}{|1-e^{-2\beta L} {\cal R}_\text{\tiny{L}}(k_F^+){\cal R}_\text{\tiny{R}}(k_F^+)|^2},
\end{equation}
 with $\beta$=Im($k_F$). The above expression shows that the behavior of $g_0$ as a function of $L$ is universal, as the above results apply to arbitrary periodic chains and lead potentials.
 
In the limit of very long chains, the left and right contacts decouple and the adiabatic conductance becomes
 \begin{equation}
g_0(L)= \frac{4}{\pi}\mbox{Im}[{\cal R}_\text{\tiny{L}}(k_F^+)] \mbox{Im}[{\cal R}_\text{\tiny{R}}(k_F^+)] e^{-2\beta L}.
\end{equation}
The reflection coefficients are directly proportional to the local density of states $\rho_{\epsilon_F}^\text{\tiny{L/R}}$ at the contact edges and at the Fermi energy. Indeed, according to Ref.~\onlinecite{Prodan:2006uq},
\begin{equation}
\mbox{Im}[{\cal R}_\text{\tiny{L/R}}(k_F^+)] =\frac{d\epsilon_F / d\beta}{\psi_{k_F}(0)^2} \rho_{\epsilon_F}(x)|_{x=\mp L/2},
\end{equation}
which allows us to rewrite the adiabatic conductance as
\begin{equation}
g_0(L)=  \frac{4}{\pi} \left [\frac{d\epsilon_F / d \beta}{\psi_{k_F}(0)^2} \right ]^2
\rho_{\epsilon_F}^\text{\tiny{L}} \rho_{\epsilon_F}^\text{\tiny{R}} e^{-2\beta L}.
\end{equation}
In the next section, the above expression will be generalized to linear molecular chains in 3D.

 \section{Molecular Chains in 3D}
 
 Real molecular or atomic chains are not strictly one dimensional, even though they are electronic systems highly confined in two dimensions. Compared to the strictly 1D systems discussed in the previous section, the fundamental difference is that a 3D chain has an infinite number of linearly independent Bloch solutions at any given energy, rather than just two. In one dimension we have scattering processes only between $k$ and $-k$ while in 3D chains we have scattering processes among an infinite set of wavenumbers.

\begin{figure}
  \includegraphics[width=8cm]{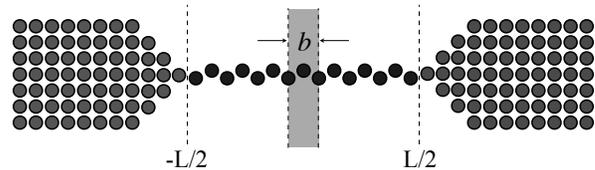}\\
  \caption{Schematic of the 3D chain+leads. The green shading indicates the region that is periodically repeated in the construction of the periodic potential $V_0$.}
  \label{3dchain}
\end{figure} 

In the limit of very long chains, like the one schematically shown in Fig.~\ref{3dchain}, we can decompose the effective potential $V_{\text{\tiny{eff}}}$ of the chain+leads into a perfectly periodic piece, $V_0$, extending from $-\infty$ to $+\infty$, and a difference $\Delta V=V_{\text{eff}}-V_0$.  The periodic potential $V_0$ is constructed by periodically repeating the effective potential between $-b/2$ and $b/2$ ($b$ being the chain's lattice constant) in the middle of the chain (see Fig.~\ref{3dchain}). For very long chains, the periodic potential constructed in this way should converge to the bulk effective potential of the chain rigidly shifted by an amount required to align the Fermi levels of the chain and the leads. The entire procedure is illustrated with a concrete example in Fig.~\ref{deco}.

We may regard the effective Hamiltonian of the chain + leads as a periodic effective Hamiltonian,
 \begin{equation}
 H_0=-\frac{1}{2}\nabla^2+V_0(\br), \ V_0(\br + b{\bf e}_z)=V_0(\br),
 \end{equation}
strongly perturbed by the potential $\Delta V$. The Kohn-Sham Hamiltonian of the entire system is then
 \begin{equation}
 H=H_0+\Delta V_L(\br)+\Delta V_R(\br),
 \end{equation}
where we divided  $\Delta V$ into left and right parts. We are not making here the simplification that $\Delta V_{L,R}$ be zero for $-L/2<z<L/2$, but we assume  that $\Delta V_{L,R}$ do decay fast to zero as we move away from the contacts. That is exactly the case in the numerical example shown in Fig.~\ref{deco}.
  
 \subsection{Green's function for the unperturbed chain}
 
 The key result that allows us to make analytical progress in evaluating the conductivity tensor, is a compact expression for $G_\epsilon^0$=$(\epsilon-H_0)^{-1}$. This expression is a generalization of Eq.~(\ref{compactGreen}) and was recently derived in Ref.~\onlinecite{Prodan:2006yq}. In the same reference, it was shown that the multitude of Bloch functions, $\psi_{n,k}$, and their corresponding energies, $\epsilon_{n,k}$, 
 \begin{equation}
 \begin{array}{c}
 [-\frac{1}{2}\nabla^2 + V_0(\br)]\psi_{n,k}(\br)=\epsilon_{n,k}\psi_{n,k}(\br) \medskip \\
  \psi_{n,k}(\br+b{\bf e}_z)=e^{ikb}\psi_{n,k}(\br),
 \end{array}
 \end{equation}
are in fact different branches of global, multi-valued functions  $\psi_{k}$ and $\epsilon_k$, defined on a suitable Riemann surface. Later on, we will explicitly calculate this surface for a simple system. The band indices do not have any meaning for complex values of $k$, since the bands touch and hybridize as we move into the complex $k$-plane. Thus, we abandon the band index and retain only the $k$ index, being understood that $k$  lives on a Rimemann surface.

The Green's function of the infinite molecular chain is given by:\cite{Prodan:2006yq}
 \begin{equation}\label{green0}
 G^0_\epsilon(\br,\br')=\sum_i\frac{\psi_{-k_i}(\br_<)\psi_{k_i}(\br_>)}{i\partial_{k}\epsilon_{k_i}},
 \end{equation}
where $\{k_i\}$ is the infinite sequence of wavenumbers on the Riemann surface such that $\epsilon_{k_i}=\epsilon$ and $\br_<  /\br_>= \br / \br'$ if $z < z'$ and $\br_<  /\br_>= \br'/ \br$ otherwise.  The Bloch functions are normalized so that
 \begin{equation}
\frac{1}{b} \int \limits _{0<z<b} \psi_k(\br)\psi_{-k}(\br) =1,
 \end{equation}
and their phase is fixed as before.
 
\subsection{Green's function for the chain+leads} 

 We calculate the Green's function for the chain+leads from
 \begin{equation}\label{greenf}
 G_\epsilon=G_\epsilon^0+G_\epsilon^0 T_\epsilon G_\epsilon^0,
 \end{equation}
 where the $T_\epsilon$ matrix is given by
 \begin{equation}\label{tmatrix}
 T_\epsilon=\Delta V+\Delta V G_\epsilon \Delta V.
 \end{equation}
 We can naturally decompose the $T$ matrix as
 \begin{equation}
 T_\epsilon=T_\text{\tiny{L}}+T_\text{\tiny{R}}+T_\text{\tiny{LR}}+T_\text{\tiny{RL}}.
 \end{equation}
Taking $\br$ and $\br'$ deep inside the chain, we obtain
 \begin{equation}\label{green}
 \begin{array}{c}
 G_\epsilon(\br,\br')=G_\epsilon^0(\br,\br')-\sum\limits_{i,j} \frac{1 }{\partial_k \epsilon_{k_i} \partial_k \epsilon_{k_j}} \times \medskip \\
 \left \{T_\text{\tiny{L}}^{ij}\psi_{k_i}(\br)\psi_{k_j}(\br')+T_\text{\tiny{R}}^{ij}\psi_{-k_i}(\br)\psi_{-k_j}(\br') \right . \medskip \\
\left . +T_\text{\tiny{LR}}^{ij}\psi_{k_i}(\br)\psi_{-k_j}(\br')+T_\text{\tiny{RL}}^{ij}\psi_{-k_i}(\br)\psi_{k_j}(\br') \right \}
 \end{array}
\end{equation}
 where
 \begin{equation}\label{matrixt}
 \begin{array}{ll}
 T_\text{\tiny{L}}^{ij} = \langle \psi_{-k_i}^\ast|T_\text{\tiny{L}}|\psi_{-k_j} \rangle,&    T_\text{\tiny{R}}^{ij} = \langle \psi_{k_i}^\ast|T_\text{\tiny{R}}|\psi_{k_j} \rangle \medskip \\
 T_\text{\tiny{LR}}^{ij} = \langle \psi_{-k_i}^\ast|T_\text{\tiny{LR}}|\psi_{k_j} \rangle,&  T_\text{\tiny{RL}}^{ij} = \langle \psi_{k_i}^\ast|T_\text{\tiny{RL}}|\psi_{-k_j} \rangle.
 \end{array}
 \end{equation}
 These coefficients depend, of course, on the energy $\epsilon$. 
 
At this point we have obtained the exact dependence of $G_\epsilon$ on the coordinates $\br$ and $\br'$, which is essential for evaluating the conductivity tensor. However, the expression of the adiabatic conductance in Eq.~(\ref{adiabaticg}), with the conductivity tensor given in Eq.~(\ref{basicsigma}), is far too complex to be approached directly. The next subsection is another key step in our calculation, which introduces a generalized Wronskian with some remarkable properties that are essential to overcoming the difficulty. 
  
 \subsection{Generalized Wronskian}
 
One can generalize the 1D Wronskian in several ways. Here we choose the following  definition:
 \begin{equation}
 W(\psi,\phi)=\int d \br_\bot  \ \psi(\br_\bot,z) \overleftrightarrow{\partial_z} \phi(\br_\bot,z).
 \end{equation}
This Wronskian appears when calculating the adiabatic conductance in Eq.~(\ref{adiabaticg}), using the conductivity tensor of Eq.~(\ref{basicsigma}) and the above expression for the Green's function of the chain+leads. The calculation of the conductance further simplifies by virtue of the following remarkable property:
\begin{equation}\label{wrproperty}
\left \{
\begin{array}{l}
W(\psi_{k_i},\psi_{k_j})=0 \medskip \\
W(\psi_{k_i},\psi_{-k_j})=-2i\partial_k \epsilon_{k_i}\delta_{k_i,k_j},
\end{array}
\right .
\end{equation}
where $\{k_i\}$ is the sequence of wavenumbers corresponding to an arbitrary complex energy $\epsilon$.

To prove Eq.~(\ref{wrproperty}), we consider two arbitrary Bloch functions,
\begin{equation}
\begin{array}{l}
[-\frac{1}{2}\nabla^2+V_0]\psi_{k}=\epsilon_k \psi_k \medskip \\

[-\frac{1}{2}\nabla^2+V_0]\psi_{k'}=\epsilon_{k'} \psi_{k'}.
 \end{array}
\end{equation}
From the two Schroedinger equations we obtain the following straightforward identity
\begin{equation}
\begin{array}{c}
\int \limits _{z_1<z<z_2} d\br \ [\psi_k \nabla^2\psi_{k'}-\psi_{k'}\nabla^2\psi_{k}] \medskip \\
=2(\epsilon_k-\epsilon_{k'})\int \limits _{z_1<z<z_2} d\br \ \psi_{k}\psi_{k'},
\end{array}
\end{equation}
where $z_1$ and $z_2$ are arbitrary. The integrand in the left hand side of the above equation can be expressed as the divergence of a vector field. Furthermore, an integration by parts leads to
 \begin{equation}\label{parts}
\begin{array}{c}
\int d {\bf x}_\bot \left \{ [\psi_k \overleftrightarrow{\partial_z} \psi_{k'}]_{z=z_2}  
-[ \psi_{k} \overleftrightarrow{\partial_z} \psi_{k'}]_{z=z_1} \right \} \medskip \\
=2(\epsilon_k-\epsilon_{k'})\int \limits_{z_1<z<z_2} d\br \ \psi_{k}\psi_{k'}.
\end{array}
\end{equation}
 We can immediately see that, if $\epsilon_k=\epsilon_{k'}$, then
 \begin{equation}
 W(\psi_{k},\psi_{k'}) _{z=z_2}=W(\psi_{k},\psi_{k'}) _{z=z_1},
 \end{equation}
 i.e. the Wronskian of the two Bloch functions is independent of $z$.
Now we choose $z_1=0$ and $z_2=b$ in Eq.~(\ref{parts}). Using the fundamental property of the Bloch functions, we finally obtain
 \begin{equation}
 \begin{array}{c}
 [e^{i(k+k')b}-1] \int d {\bf x}_\bot [\psi_k \overleftrightarrow{\partial}_z \psi_{k'}]_{z=0}  \medskip \\
 =2(\epsilon_k-\epsilon_{k'})\int \limits_{0<z<b} d \br \ \psi_{k}\psi_{k'}.
\end{array}
\end{equation}
Thus, if $\epsilon_k=\epsilon_{k'}$ but $k+k' \neq 0$, the Wronskian of the two Bloch functions is zero. Furthermore, by taking the limit $k'\rightarrow -k$ we obtain exactly the second line of Eq.~(\ref{wrproperty}).
 
\subsection{Adiabatic conductance}

{\it Insulating chains.} To compute the adiabatic conductance, we insert the Green's function (Eq.~\ref{green}) into the expression for the conductivity tensor (Eq.~\ref{basicsigma}) and compute Eq.~(\ref{adiabaticg}). This process generates Wronskians that need to be carefully counted. In addition, we need to be careful when going from $\epsilon_F^+$ to $\epsilon_F^-$. 

Let $\{k_i\}$ be the sequence of wavenumbers corresponding to $\epsilon_F^+$. Since $\epsilon_F$ falls in an energy gap, all wavenumbers are complex. When passing to $\epsilon_F^-$, the wavenumbers remain the same, but the $T$ matrix is different. This is because we cross the continuum spectrum of the leads. When $T$ is evaluated at $\epsilon_F^-$ we call it $\tilde{T}$. With this notation, after counting the Wronskians, the exact expression for the adiabatic conductance is
\begin{eqnarray}\label{exactg}
& &g_0 = -\frac{1}{\pi} \sum_i \frac{T_\text{\tiny{LR}}^{ii}+\tilde{T}_\text{\tiny{LR}}^{ii}}{i\partial_k \epsilon_{k_i} } \\
&-& \frac{1}{\pi}\sum_{i,j} \frac{T_\text{\tiny{L}}^{ij}\tilde{T}_\text{\tiny{R}}^{ij}+\tilde{T}_\text{\tiny{L}}^{ij} T_\text{\tiny{R}}^{ij} - T^{ij}_\text{\tiny{LR}}\tilde{T}^{ij}_\text{\tiny{RL}} - \tilde{T}^{ij}_\text{\tiny{LR}} T^{ij}_\text{\tiny{RL}} }
 {\partial_k \epsilon_{k_i} \partial_k \epsilon_{k_j}}. \nonumber
 \end{eqnarray}
 No approximation has been made so far.

We now show that Eq.~(\ref{exactg}) greatly simplifies in the limit $L\rightarrow \infty$. We introduce the following notation:
 \begin{equation}
\beta_i = \text{Im}(k_i), \ \  \beta_{\text{min}} = \min_i \{\beta_i \}.
 \end{equation}
 First we demonstrate that:
 \begin{equation}\label{step1}
 T_\text{\tiny{LR}}=[1+o(e^{-2\beta_{\text{min}} L})]T_\text{\tiny{L}} G_\epsilon^0 T_\text{\tiny{R}}.
 \end{equation}
This follows from the following considerations. First, we notice that every time that $G_\epsilon^0$ is sandwiched between a $\Delta V_\text{\tiny{L}}$ and a $\Delta V_\text{\tiny{R}}$, it becomes of order $o(e^{-\beta_{\text{min}}L})$. This follows from the exact expression of $G_\epsilon^0$ given above. Next, we iterate Eqs.~(\ref{greenf}) and (\ref{tmatrix}) to generate a formal series for $T_{LR}$:
\begin{equation}\label{expansion}
\begin{array}{c}
T_\text{\tiny{LR}}=\Delta V_\text{\tiny{L}} G_\epsilon^0 \Delta V_\text{\tiny{R}} +\Delta V_\text{\tiny{L}} G_\epsilon^0 \Delta V G_\epsilon^0 \Delta V_\text{\tiny{R}} \medskip \\
+ \Delta V_\text{\tiny{L}} G_\epsilon^0 \Delta V G_\epsilon^0 \Delta V G_\epsilon^0 \Delta V_\text{\tiny{R}} + \ldots
\end{array}
\end{equation}
Let us consider, for instance, the third term in Eq.~(\ref{expansion}):
\begin{equation}\label{third}
\begin{array}{c}
T_\text{\tiny{LR}}^3 =  \Delta V_\text{\tiny{L}} G_\epsilon^0 \Delta V_\text{\tiny{L}} G_\epsilon^0 \Delta V_\text{\tiny{L}} G_\epsilon^0 \Delta V_\text{\tiny{R}} \medskip \\
 +\Delta V_\text{\tiny{L}} G_\epsilon^0 \Delta V_\text{\tiny{L}} G_\epsilon^0 \Delta V_\text{\tiny{R}} G_\epsilon^0 \Delta V_\text{\tiny{R}} \medskip \\
  +\Delta V_\text{\tiny{L}} G_\epsilon^0 \Delta V_\text{\tiny{R}} G_\epsilon^0 \Delta V_\text{\tiny{R}} G_\epsilon^0 \Delta V_\text{\tiny{R}} \medskip \\
   +\Delta V_\text{\tiny{L}} G_\epsilon^0 \Delta V_\text{\tiny{R}} G_\epsilon^0 \Delta V_\text{\tiny{L}} G_\epsilon^0 \Delta V_\text{\tiny{R}}. \medskip \\
 \end{array}
 \end{equation}
In the first three terms, $G_\epsilon^0$ is sandwiched between a $\Delta V_\text{\tiny{L}}$ and a $\Delta V_\text{\tiny{R}}$ only once, whereas all the three $G_\epsilon^0$ that appear in the last term are sandwiched between a $\Delta V_\text{\tiny{L}}$ and $\Delta V_\text{\tiny{R}}$. As a consequence, the ratio of the fourth term and anyone of the first three terms is of order $o(e^{-2\beta_{\text{min}} L})$. By applying this argument to all the terms in Eq.~(\ref{expansion}) we obtain:
\begin{equation}\label{nterm}
\begin{array}{cccc}
T_\text{\tiny{LR}}^n=\sum\limits_{k=1}^n & \underbrace{ \Delta V_\text{\tiny{L}} G_\epsilon^0  \ldots \Delta V_\text{\tiny{L}}}&  G_\epsilon^0 &\underbrace{\Delta V_\text{\tiny{R}} \ldots   G_\epsilon^0 \Delta V_\text{\tiny{R}}} \\
&k& &n+1-k
\end{array}
\end{equation}
plus terms $o(e^{-2\beta_{\text{min}}L})$ times smaller. 

Next, we consider the expansion of $T_\text{\tiny{L}} G_\epsilon^0 T_\text{\tiny{R}} $ in powers of $G_\epsilon^0$. By applying the same arguments that led us to Eq.~(\ref{nterm}), we find that the $n$-th term in the exapnsion is equal to the right hand side of Eq.~(\ref{nterm})  plus terms $o(e^{-2\beta_{\text{min}}L})$ times smaller. This proves Eq.~(\ref{step1}). Taking the matrix elements of this equation and using the representation of the Green's function $G_\epsilon^0$ given in Eq.~(\ref{green0}) gives:
\begin{equation}\label{tlrassymptotic}
T_\text{\tiny{LR}}^{ij}=[1+o(e^{-2\beta_{\text{min}}L})]\sum_m \frac{ T_\text{\tiny{L}}^{im} T_\text{\tiny{R}}^{mj}}{i\partial_k \epsilon_{k_m}}.
\end{equation}
Similar conclusions holds for $T_\text{\tiny{RL}}^{ij}$ and for the tilde counterparts.

As shown below in Eqs.~(\ref{T1}) and (\ref{T2}), the matrix elements  $T_\text{\tiny{L}}^{ij}$ and $T_\text{\tiny{R}}^{ij}$ are exponentially small for long chains. Thus, if we retain only the leading terms, Eq.~(\ref{exactg}) becomes:
\begin{equation}\label{simpleg0}
g_0 =  \frac{1}{\pi}\sum_{i,j} \frac{(T_\text{\tiny{L}}^{ij}-\tilde{T}_\text{\tiny{L}}^{ij})(T_\text{\tiny{R}}^{ij}-\tilde{T}_\text{\tiny{R}}^{ij})}
 {\partial_k \epsilon_{k_i} \partial_k \epsilon_{k_j}} .
 \end{equation}
This is the exact asymptotic form of $g_0$. The terms that we have neglected are $o(e^{-2\beta_{\text{min}}L})$ times smaller.

Finally, we show that the matrix elements of $T$ have simple and intuitive expressions. For instance, the first factor in the numerator on the right hand side of Eq.~(\ref{simpleg0}) is:
\begin{equation}
T_\text{\tiny{L}}^{ij}-\tilde{T}_\text{\tiny{L}}^{ij}=\langle \psi_{-k_i}|\Delta V_\text{\tiny{L}} (G_{\epsilon_F^+}-G_{\epsilon_F^-})\Delta V_\text{\tiny{L}}  |\psi_{-k_j} \rangle.
\end{equation}
This can be expressed in terms of the spectral operator $\rho_{\epsilon_F}$:
\begin{equation}
\rho_{\epsilon_F}=\frac{1}{2\pi i} \left (G_{\epsilon_F^+}-G_{\epsilon_F^- } \right ).
\end{equation}
The diagonal elements, $\rho_{\epsilon_F}(x,x)$, of the spectral operator give the local density of states. By writing $\psi_k(\br)=u_k(\br)e^{ikz}$, with $u_k$ periodic, $u_k(\br+b{\bf e}_z)=u_k(\br)$, we obtain
\begin{equation}\label{T1}
T_\text{\tiny{L}}^{ij}-\tilde{T}_\text{\tiny{L}}^{ij}=e^{\frac{i}{2}(k_i+k_j) L}\Theta_\text{\tiny{L}}^{ij}
\end{equation}
with
\begin{equation}\label{theta1}
\begin{array}{c}
\Theta_\text{\tiny{L}}^{ij}=2\pi i \int d \br \int d \br'  e^{-i(k_i z+k_j z')} \times \medskip \\
u_{-k_i}(\br)\Delta V_\text{\tiny{L}}(\br)\rho_{\epsilon_F}(\br,\br')\Delta V_\text{\tiny{L}}(\br')u_{-k_i}(\br'),
\end{array}
\end{equation}
where $\br$ and $\br'$ are measured from the left end of the chain, $z=-L/2$. Similarly
\begin{equation}\label{T2}
T_\text{\tiny{R}}^{ij}-\tilde{T}_\text{\tiny{R}}^{ij}=e^{\frac{i}{2}(k_i+k_j) L}\Theta_\text{\tiny{R}}^{ij}
\end{equation}
with
\begin{equation}\label{theta2}
\begin{array}{c}
\Theta_\text{\tiny{R}}^{ij}=2\pi i \int d \br \int d \br'  e^{i(k_i z+k_j z')} \times  \medskip \\
u_{k_i}(\br)\Delta V_\text{\tiny{R}}(\br)\rho_{\epsilon_F}(\br,\br')\Delta V_\text{\tiny{R}}(\br')u_{k_i}(\br'),
\end{array}
\end{equation}
where $\br$ and $\br'$ are measured from the right end of the chain, $z=L/2$. In the limit $L\rightarrow \infty$, the $\Theta$ coefficients become independent of $L$. By inserting Eqs.~(\ref{T1}) and (\ref{T2}) into Eq.~(\ref{simpleg0}), we obtain the following simple asymptotic form for $g_0$:
\begin{equation}\label{insulatingg0}
g_0(L)=\frac{1}{\pi}\sum\limits_{i,j} \frac{\Theta_\text{\tiny{L}}^{ij} \Theta_\text{\tiny{R}}^{ij}}{\partial_k \epsilon_{k_i} \partial_k \epsilon_{k_j}}  e^{i(k_i+k_j)L}.
\end{equation}
We notice that the integrands in Eqs.~(\ref{theta1}) and (\ref{theta2}) contain the following terms:
\begin{equation}
\begin{array}{l}
e^{-i(k_i z+k_j z')}\Delta V_\text{\tiny{L}}(\br)\Delta V_\text{\tiny{L}}(\br') \medskip \\  
e^{i(k_i z+k_j z')} \Delta V_\text{\tiny{R}}(\br)\Delta V_\text{\tiny{R}}(\br'),
\end{array}
\end{equation}
which are highly localized near the left and the right contacts, respectively. This shows that the conductance of the molecular device is only determined by the properties of the chain and of the contacts.

Strictly speaking, the asymptotic form of $g_0(L)$ is determined by the wavenumber $k$ such that $\text{Im}(k) =\beta_{\text{min}}$. However, for complex molecular chains such as carbon nanotubes,\cite{Pomorski:2004nx} there may be many wavenumbers with similar imaginary parts, especially when the valence and the conduction bands are highly degenerate.

{\it Metallic chains.} As in the insulating case, let us consider the infinite sequence of wavenumbers $\{k_i\}$ for which $\epsilon_{k_i} = \epsilon_{F}$. Since the Fermi energy falls within allowed bands, some of the $k_i$ are real. We can restrict ourselves to the latter, because complex wavenumbers lead to exponentially small contributions to $g_0$ in the limit $L\rightarrow \infty$.

In order to compute the conductivity tensor in Eq.~(\ref{basicsigma}) we need both $G_{\epsilon_F^+}$ and $G_{\epsilon_F^-}$. When going from $\epsilon_F^+$ to $\epsilon_F^-$, $k_i$ becomes $-k_i$. Thus:
\begin{equation}
\ T^{ij} \rightarrow \tilde{T}^{ij}=(T^{ij})^\ast, \ \partial_k \epsilon_{k_i} \rightarrow -\partial_k \epsilon_{k_i}.
\end{equation}
Following the same path as in the insulating case, we obtain:
\begin{eqnarray}\label{gmetal}
& &g_0=\frac{1}{\pi}\sum_i \left [ 1+\frac{2\text{Im} T_\text{\tiny{LR}}^{ii}}{\partial_k \epsilon_{k_i}} \right ] \\
&-&\frac{1}{\pi}\sum_{ij} \frac{ |T_\text{\tiny{L}}^{ij}|^2+|T_\text{\tiny{R}}^{ij}|^2-|T_\text{\tiny{LR}}^{ij}|^2-|T_\text{\tiny{RL}}^{ij}|^2 }
{\partial_k \epsilon_{k_i} \partial_k \epsilon_{k_j}}. \nonumber
\end{eqnarray}
At zero temperature, the left and right contacts do not decouple and all the terms in Eq.~(\ref{gmetal}) should be retained, even in the limit of very long chains. However, at finite temerature, the Fermi energy acquires a small imaginary part and the same decoupling mechanism of the insulating chains holds. In this regime, Eq.~(\ref{gmetal}) simplifies to:
\begin{equation}
g_0=\frac{N}{\pi}-\frac{4}{\pi}\sum_{ij}\frac{\text{Re} T_\text{\tiny{L}}^{ij} \text{Re} T_\text{\tiny{R}}^{ij} }{\partial_k \epsilon_{k_i} \partial_k \epsilon_{k_j} },
\end{equation}
where $N$ is the number of real $k$ points such that $\epsilon_k=\epsilon_F$. 

If the potential $V_0(\br)$ is separable in longitudinal and transverse coordinates, $V_0(\br)=V_0^{\text{\tiny{1}}}(x,y)+V_0^{\text{\tiny{2}}}(z)$, $N$ is simply the number of transverse modes.\cite{A.-Kamenev:2001uq} In the general case, $N$ is the number of eigenstates of the $T$ matrix at $\epsilon = \epsilon_F$. In our approach, $N$ can be extracted from the band structure of the chain. For instance, with reference to the band structure in Fig.~\ref{bands}, if $\epsilon_F$ falls within the lowest energy band,  $N=1$. If $\epsilon_F$ falls within the third band (in order of increasing energy) and crosses the band at more than two points, $N=2$.

The matrix elements of $T$ in Eq.~(\ref{gmetal}) exhibit oscillatory behavior with $L$. Thus the same conclusion that we reached for strictly 1D case remains valid, namely, $g_0$ should exhibit oscillatory behavior with $L$. Again, the phase of the oscillations will be affected by the resonant behavior of the $T$ matrix.

  \begin{figure}
 \includegraphics[width=8cm]{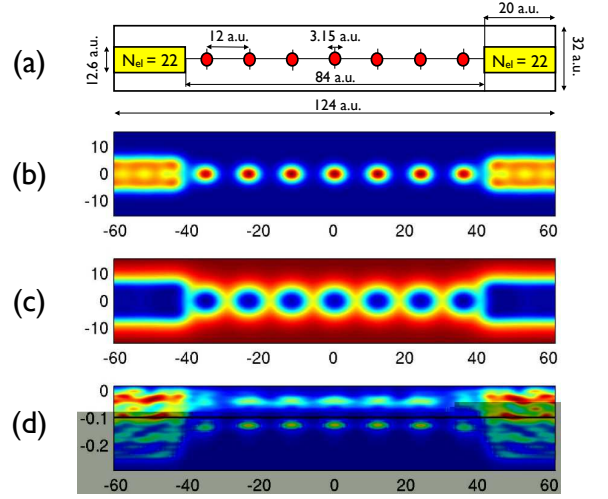}\\
 \caption{Upper panel: The self consistent electron density. Middle panel: The self consistent potential. Lower panel: The local density of states, integrated over the xy cooridnats. The black line shows the position of the Fermi level. The numbers represent atomic units.}
\label{equilibrium}
\end{figure} 

\section{Numerical example}

In this section we provide an example that illustrates how the above ideas can be implemented numerically. We consider an insulating chain made of quantum dots (qdots). These are jellium qdots modeled by a spherically symmetric potential well:
\begin{equation}
V_q (\br)= \frac{-v_0}{1+\exp[(r-r_0)/\xi)}, \ \ v_0 = 0.1 \text{a.u.}
\end{equation}
The smoothness of the potential edge is controlled by $\xi$, which we take here to be $\xi=1$ a.u.. We also added a neutralizing background of positive charge, with charge density:
\begin{equation}
n_+ (\br)= \frac{n_0}{1+\exp[(r-r_0)/\xi)},
\end{equation}
where $n_0$ has $r_s = 2.5$ a.u. and $r_0 = 3.15$ a.u.. These values were chosen so that $n_+$ integrates to 2e. Thus, the qdot can accommodate 2 electrons and has a complete shell. A linear periodic array of such qdots will be insulating if the distance $b$ between adjacent qdots is large enough. This condition is met by our choice $b=12$ a.u..

  \begin{figure}
 \includegraphics[width=8cm]{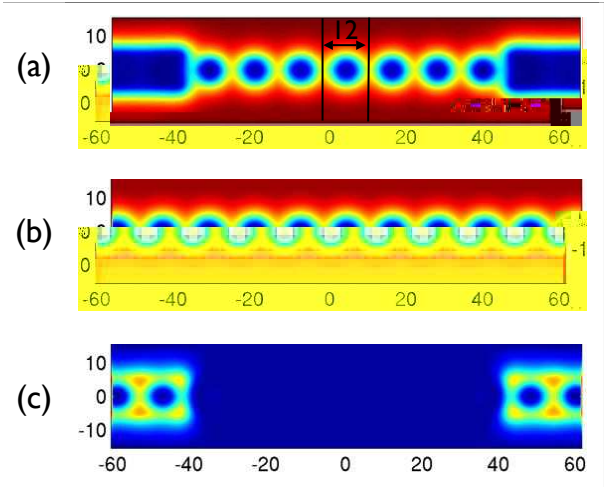}\\
 \caption{Upper panel: The effective Kohn-Sham potential of the chain + leads and the region that is repeated periodically. Middle panel: The periodic potential obtained by repeating the central piece of the upper panel. Lower panel: The difference between the effective and periodic potential. The numbers represent atomic units.}
\label{deco}
\end{figure} 

We have considered a chain of seven qdots attached to cylindrical jellium leads at both ends. The setup is shown in Fig.~\ref{equilibrium}(a). Periodic boundary conditions were adopted along $z$ (with a cell parameter of 124 a.u.), while free boundary condition were used in the lateral directions.

In Figs.~\ref{equilibrium}(b) and (c) we display the self-consistent density and potential $V_{\text{\tiny{eff}}}(\br)$, respectively. Fig.~\ref{equilibrium}(d) shows the local density of states integrated over the xy coordinates. Here one can see the bands of the chain and the Fermi level, which is located between the first two bands.

Fig.~\ref{deco} illustrates the procedure that we use to construct the periodic potential  $V_0(\br)$. This is obtained by repeating infinite times the part of $V_{\text{\tiny{eff}}}(\br)$ comprised between $-b/2$ and $b/2$. The resulting periodic potential is shown in Fig.~\ref{deco}(b). Fig.~\ref{deco}(c) shows the difference $\Delta V = V_{\text{\tiny{eff}}}-V_0$. The potential is cylindrically symmetric, so the $L_z$ quantum number is conserved. Since the lowest bands correspond to the $L_z=0$ sector, the entire discussion will be restricted to this sector.

\begin{figure}
 \includegraphics[width=8cm]{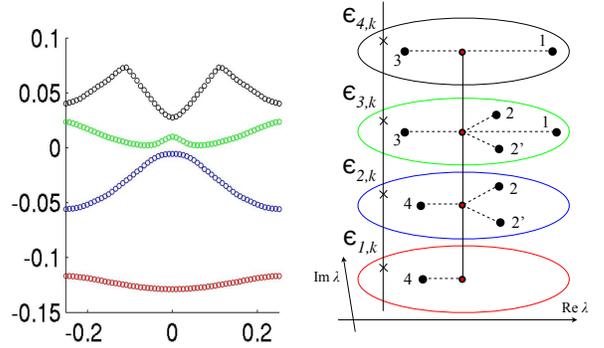}\\
 \caption{Left panel: The lowest bands of the periodic potenial. Right panel: The Riemann surface corresponding to these bands.}
\label{RiemannS}
\end{figure}

The next step is the complex band structure calculation and the construction of the Riemann surface for the periodic potential $V_0(\br)$. The theory of the complex band structure for a general periodic potential was discussed in Ref.~\onlinecite{Prodan:2006yq}. Based on this theory we compute the Riemann surface of the bands for $V_0(\br)$. In Fig.~\ref{RiemannS} we show the four lowest bands and the corresponding section of the Riemann surface, which consists of four unit disks. The Riemann surface is represented using the natural variable $\lambda = e^{ikb}$. The full dots indicate branch points. The dashed lines connecting the branch points to the center of the disks represent branch cuts, which always occur in pairs. In Fig.~\ref{RiemannS} the pairs are located on adjacent disks. The disks become connected via these branch cuts.

In order to construct the Riemann surface, we diagonalize  the $k \cdot p$ Hamiltonian,
\begin{equation}
H_k = -\frac{1}{2}\nabla_\bot^2-\frac{1}{2} (-i\partial_z + k)^2 + V_\text{\tiny{per}},
\end{equation}
defined in the interval $[-b/2,b/2]$ with periodic boundary conditions. The $k \cdot p$ Hamiltonian was implemented on a real space grid, using a 5 point finite difference representation for the second derivatives. The grid spacing was 0.5 a.u.. 

\begin{figure}
 \includegraphics[width=8.6cm]{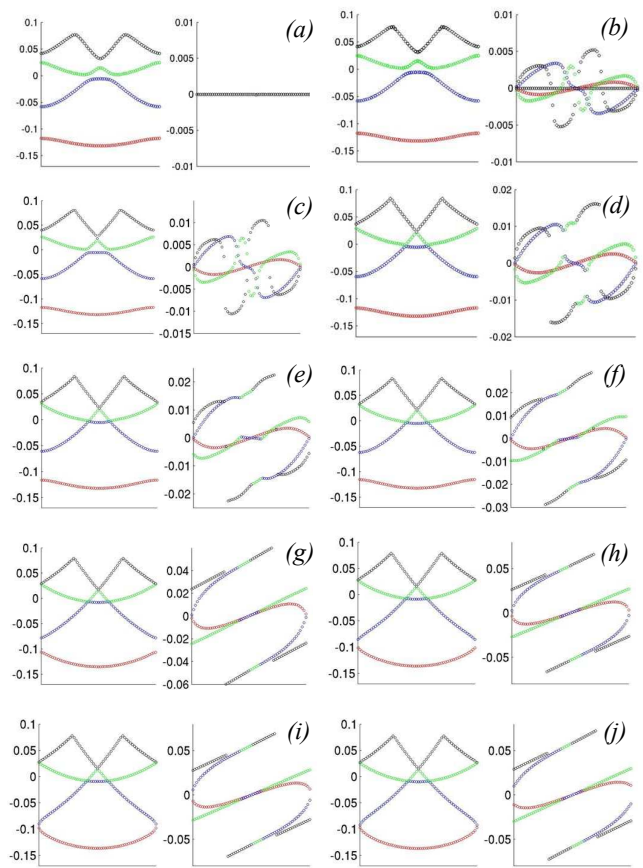}\\
 \caption{Complex band calculations. Each panel shows the real (left) and imaginary (right) parts of $\epsilon_{n,k}$, $n=1,\ldots,4$, for different values of $\text{Im} k$. In the panels from (a) to (f) Im($k$) was uniformly increased from 0 to 0.06 (in a.u.). In the panels from (g) to (j), Im($k$) = 0.1, 0.11, 0.115 and 0.12 (a.u.), respectively.}
\label{bands}
\end{figure}

Each panel of Fig.~\ref{bands} shows the eigenvalues $\epsilon_{n,k}$, $n=1,\ldots,4$ of the $k \cdot p$ Hamiltonian as functions of $k$, when the imaginary part of $k$ is fixed and the real part is varied from $-\pi/b$ to $\pi/b$. When $k$ moves on this line in the complex $k$-plane, the variable $\lambda = e^{ikb}$ moves on a circle; the larger Im($k$) the smaller the radius of this circle. We report in Fig.~\ref{bands} a set of panels going from (a) to (j). Each panel corresponds to a different value of $\text{Im}(k)$, increasing from $\text{Im}(k)=0$ a.u. (panel (a)) to $\text{Im}(k)=0.12$ a.u. (panel (j)). Each panel shows two diagrams: the one on the left represents Re($\epsilon_{n,k}$) vs Re$(k)$, and the one on the right represents Im($\epsilon_{n,k}$) vs Re$(k)$.

In order to understand the analytic structure of the bands, for a given value of $k$ (or $\lambda$), one may draw a vertical line like the one in the right panel of Fig.~\ref{RiemannS}. This vertical line intersects the Riemann surface infinite times. The infinite sequence of complex energies $\epsilon_{n,k}$, $n=1,\ldots,\infty$, of the $ k \cdot p$ Hamiltonian are in fact the values of the same function, $\epsilon_k$, evaluated at these intersection points. If we move $\lambda$ on a circle of radius $e^{-b\text{Im}(k)}$, these intersection points generate certain contours on the Riemann surface. When $\epsilon_{k}$ is evaluated for all $\lambda$ located on these contours, one obtains certain paths in the complex-energy plane. The complex bands of Fig.~\ref{bands} are in fact examples of such paths.

If a contour remains on the same Riemann sheet as $\lambda$ completes the circle, e.g. the n-th sheet, then $\epsilon_{n,k}$ moves continuously on a closed path in the complex energy plane. In other words, both Re($\epsilon_{n,k}$) and Im($\epsilon_{n,k}$) return to their initial value upon completing the circle. If this does not happen, i.e. the complex $\epsilon_{n,k}$ does not return to its initial value, the contour has intersected a branch cut. When $\lambda$ hits a branch point connecting the sheets $n$ and $m$, the complex bands generated by $\epsilon_{n,k}$ and $\epsilon_{m,k}$ intersect.

The Riemann surface of Fig.~\ref{RiemannS} was computed as follows.  For small Im($k$), all $\epsilon_{n,k}$ return to their initial value when Re($k$) hits the right edge of the Brillouin zone. This is what we see in the first two panels of Fig.~\ref{bands}. As we increase Im($k$), we hit the first branch point (labeled 1 in Fig.~\ref{RiemannS}). This branch point connects the third and the fourth sheet, thus the real and the imaginary parts of  $\epsilon_{3,k}$ and $\epsilon_{4,k}$ are equal at this particular value of $k$. This is exactly what we see in panel (c). By further increasing Im($k$), we hit the second set of branch points, indicated by 2 and 2' in Fig.~\ref{RiemannS}. These branch points connect the second and the third sheet, thus the real  and imaginary parts of $\epsilon_{2,k}$ and $\epsilon_{3,k}$ are equal at these $k$ points. In panel (d) we have crossed these branch points. Indeed, by examining the real and imaginary parts of $\epsilon_{2,k}$ and $\epsilon_{3,k}$, we see that they no longer remain on the same sheet. For example, $\epsilon_{2,k}$ moves from the second to the third sheet and then to the fourth sheet. Next we hit the third branch point, and we see this happening in panels (d) and (e). At last, we hit the fourth branch point, which connects the first and second sheet. We see this happening in panels (i) and (j). By further increasing Im($k$) we do not find any additional branch points on the first three Riemann sheets. On the fourth sheet, there are additional branch points connecting this sheet with upper sheets.

Finally, we compute the $\Theta$ coefficients. In the case that we are considering here, the Fermi level falls between the first and the second band. The complex band calculations provide the corresponding sequence of $\{k_i\}$ wavenumbers. We need to consider only the $k$ wavenumber located on the branch cut connecting the first and the second Riemann sheets. Consequently, we can drop the $ij$ indices in Eqs.~(\ref{theta1}) and (\ref{theta2}). The complex band calculations also provide the left/right eigenvectors $u_{-k}$ and $u_{k}$ that enters the definition of the $\Theta$ coefficients. The coefficients are then calculated as follows. Rather than computing the spectral operator, we obtain directly the functions $G_{\epsilon_F^\pm} \Delta V_\text{\tiny{L/R}} \psi_{\mp k}$, by solving for $\Psi$ in
\begin{equation}\label{spectral}
(\epsilon_F^\pm-H)\Psi = \Delta V_\text{\tiny{L/R}} \psi_{\mp k}.
\end{equation}
Eq.~(\ref{spectral}) was solved on a real space grid with the same grid spacing as in the case of $k \cdot p$ equations. The metallic leads were increased until the supercell reached 250 a.u. in length. With this supercell, we could use an imaginary part of $\delta=0.005$ Ha for $\epsilon_F^\pm$. After we calculate $\Psi$, we take its scalar product with $2\pi i \Delta V_ \text{\tiny{L/R}} \psi_{\mp k}$ to finally get $\Theta\text{\tiny{L/R}}$ from Eqs.~(\ref{theta1}) and (\ref{theta2}).

\begin{figure}
 \includegraphics[width=8cm]{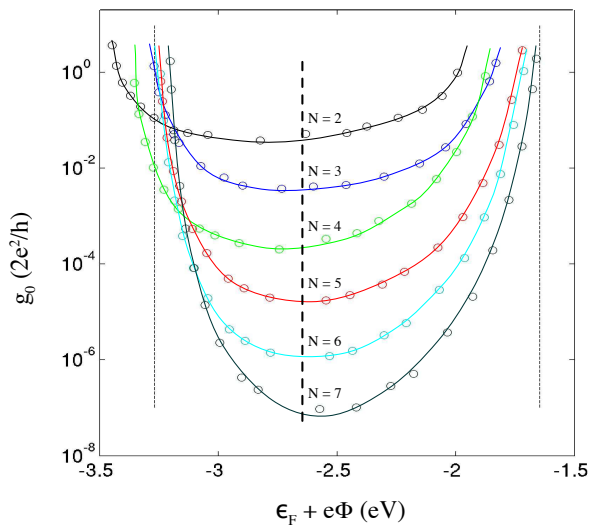}\\
 \caption{The differential conductance for chain+leads structures containing 2 (black), 3 (blue), 4 (green), 5 (red), 6 (cyan) and 7 (indigo) qdots, as function of the applied bias potential. The thick vertical dashed line represents the converged position of the Fermi level. The thin dotted lines indicate the edges of the valence and conduction bands for the structure containing 7 qdots.}
\label{co}
\end{figure}

To get more insight into the transport properties of the chain, we have rigidly moved the Fermi level by a biased potential ${\cal V}$ and calculated $g_0$ at $\epsilon_F+e\Phi$. To first approximation, this will give the non-linear differential conductance of the chain.\cite{Magoga:1997fk} We have repeated the above steps for chains containing 2, 3,\ldots, 7 qdots and the results are shown in Fig.~\ref{co}. 

\section{Discussion} 

The self-consistent potential is not the same for the different structures considered in Fig.~\ref{co}. Consequently, the periodic potential used in the calculation of $g_0$ is different from case to case. However, due to nearsightedness,\cite{Prodan:2005vn} in the middle of the chain we expect these self-consistent potentials to converge to the bulk value, as the chain's length gets larger and larger. In our calculations, we have seen a substantial change in the periodic potential for the systems containing 2, 3 and 4 qdots. However, the band energies of the periodic potentials for the structures containing 5, 6 and 7 qdots are almost the same, indicating that the periodic potential is almost converged. The edges of the valence ($\epsilon_v$) and conduction ($\epsilon_c$) bands for the periodic potential corresponding to the structure with 7 qdots are indicated by thin, vertical dotted lines in Fig.~\ref{co}.  The Fermi level, marked by a thick vertical dashed line in Fig.~\ref{co}, was practically the same for all systems.

Each circle in Fig.~\ref{co} represents the output of a single numerical run. One can see that $g_0$ varies smoothly with the bias potential. This indicates that the super-cell used in the calculation is large enough, so that the density of states appears smooth when computed with our damping parameter $\delta = 0.005$ Ha. The points in Fig.~\ref{co} can be connected by a line having the characteristic shape found in Ref.~\onlinecite{Magoga:1997fk}. There is one important difference, however: our line shape is not parabolic as in Ref.~\onlinecite{Magoga:1997fk}, but has an asymmetry since the conductance minimum occurs closer to the valence band. 

$g_0$ is primarily determined by the variation with the bias potential of Im($k$), $\partial_k \epsilon_k$ and $\psi_k$. For very long chains, the conductance minimum occurs at the maximum  value of Im($k$). In our example, the conductance minimum is achieved when $\epsilon_F+e\Phi$ is equal to the energy of the branch point, i.e. when $k$ becomes the branch point itself. This is not a universal feature, however, because the energies of the branch points are complex, in general, and cannot be equal to $\epsilon_F+e\Phi$. For example, this is the case with alkyl chains. The asymmetry of the curves in Fig.~\ref{co} is in agreement with the theoretical results of Ref.~\onlinecite{Prodan:2006uq} about the position of the branch point, which is reviewed below.

When the gap is small relative to the width, $w$, of the valence band, one can use, quite accurately, the effective mass or quadratic approximation for the band energies, in which case finding the complex $k$ vectors for all energies in the gap is trivial. In particular, the energy of the branch point is found to be in the middle of the gap. The fact that one can use the effective mass approximation to calculate the complex $k$ wavenumbers, which give the exponential decay with the chain's length of $g_0$, implies that one can also use a simple square barrier model to calculate $g_0$. In other words, if one plugs the transmission coefficient for tunneling of a particle of mass $m^*$ (the effective mass of the electrons in the chain) through a square barrier of height $\epsilon_c -\epsilon_F-e\Phi$ in the Landauer formula, one will find the correct exponential decay behavior of $g_0$ with the chain's length. 

In the narrow band limit, which is our case, the branch point strongly shifts towards the valence band and its imaginary part is found to be:\cite{Prodan:2006uq}
\begin{equation}\label{kapprox}
\text{Im}(k)=\frac{1}{b}\ln \left[ \frac{8\sqrt{-\epsilon_v}}{bw} \right ] -\frac{1}{b}.
\end{equation}
This shift leads to the asymmetry that can be seen in Fig.~\ref{co}. For example, for the chain+leads structure containing 7 qdots, Eq.~(\ref{kapprox}) gives Im($k$)=0.14 a.u., as opposed to a value of about 3 a.u. that is provided by the effective mass approximation. The exact value of Im($k$) is 0.115 a.u.. This is a concrete example of the breakdown of the effective mass approximation for chains with narrow bands.  Because of that, the picture of electrons tunneling through a square barrier is misleading in this regime.

Note that $\partial_k \epsilon_k$ is zero at the band edges. This means that our asymptotic expression for $g_0$ diverges as we move towards the gap edges. This can be corrected by including multiple-reflection terms in $T_\text{\tiny{LR}}$. Also note that $\partial_k \epsilon_k$ is infinite at the branch point. Apparently, this leads to a singularity. However, the Bloch functions, $\psi_k$ and $\psi_{-k}$, also diverge at the branch point.\cite{Prodan:2006yq} The divergences cancel exactly and $g_0$ becomes finite and smooth at the branch point. 

Besides the exponential decay, the conductance should also display oscillatory behavior with the length of the chain. This oscillatory behavior comes from the real parts of $k_i$ in Eq.~(\ref{insulatingg0}). In our example, the oscillation is absent because $k$ is at the zone boundary. The oscillation would be present if the Fermi level were located between the second and the third band of Fig.~\ref{RiemannS}.

For all six chain+leads structures, we found that the Fermi level practically coincides with the energy of the branch point. This was quite surprising, because the branch point in our calculations is not in the middle of the gap as in Refs.~\onlinecite{Tomfohr:2002oq}-\onlinecite{Picaud:2003qf}. This sends us back to Tersoff's criterion, which says that the Fermi level is pinned at the branch point.\cite{Tersoff:1984kl} The arguments leading to this conclusion are valid for cases when the metal-induced gap states extend far into the chain and are based on the assumption of local (as opposed to global) charge neutrality. In 1D, one can use an observation by Rehr and Kohn, which says that the gap states below/above the branch point originate from the valence/conduction band.\cite{Rehr:1974hc} Then, in order to maintain the local charge neutrality somewhere deep in the chain, we must occupy only the gap states from the valence band. This pins the Fermi level at the branch point. A clear division between gap states is always possible in 1D because the energy of the branch point is real. This is also true in our example but the energy of the branch points connecting the valence and conduction bands can be complex. Thus, no such clear division of the gap states is possible. Nevertheless, Tersoff's criterion seems to be confirmed by many numerical applications, including ours.  It will be interesting to see if Rehr and Kohn's observation can be generalized from 1D to real molecular chains and if a more general explanation of Tersoff's criterion can be found.

\section{Conclusions}

In conclusion, we derived a formally exact expression for the two-terminal dc conductance within Time Dependent Current-Density Functional Theory. This general result may provide a useful starting point to go beyond the adiabatic approximation in transport.

Using a compact expression for the Green's function, previously reported in Ref.~\onlinecite{Prodan:2006yq}, and a generalized Wronskian, reported here for the first time, we derived explicit analytic expressions for the adiabatic part of the conductance. For insulating chains, we found that the adiabatic conductance is proportional to the overlap between the spectral kernel $\rho_{\epsilon_F}(\br,\br ')$, the complex $k$ Bloch functions of the chain and the potential perturbation at the contacts $\Delta V_{\text{\tiny{L/R}}}$. We also found that the conductance has an exponential decay behavior with the chain's length. This behavior may be modulated by an oscillatory factor. Both the exponential decay constant and the period of the oscillations can be predicted from complex band calculations. For metallic chains, we rediscovered the oscillatory behavior of the conductance with the chain's length. In the light of the new analytic expressions, we discussed the phase differences  in these oscillations that were observed in alkali and noble metal chains.

The results of this paper may open the possibility for quantitative theoretical predictions on tunneling transport through extremely long molecular chains that cannot be treated by current {\it ab-initio} approaches. The results can be extended to finite temperatures and thus may provide a quantitative understanding of the tunneling transport experiments involving alkyl chains grown on a silicon surface,\cite{A.-Salomon:2005kx} in both thermionic and tunneling regimes. The analysis can be also generalized to the spin dependent case, useful to understanding the tunneling magneto-resistance. We think that even in this case it may be possible to obtain quantitative predictions similar to those of Ref.~\onlinecite{Heiliger:2006tg}, without going through costly {\it ab-initio} calculations.

\acknowledgements

We like to acknowledge extremely useful discussions with Giovani Vignale and Carsten Ullrich, which ultimately pointed us to the possible additional dynamical contribution to the conductance due to the transverse fields. One of the authors (EP) also likes to acknowledge the hospitality of Prof. Walter Kohn,  who helped us to better understand the microscopic origin of the physical potential drop that is usually measured by a voltmeter.  Partial support for this work was provided by the NSF-MRSEC program through the Princeton Center for Complex Materials (PCCM), grant DMR 0213706, and by  DOE through grant DE-FG02-05ER46201. One of us (E.P.) is a PCCM fellow.


\end{document}